\documentclass[aps,showpacs]{revtex4}
 
\def\<{\langle}
\def\>{\rangle}
\begin{document}
\title{general scheme for super dense coding between  multi-parties}
\author{ X. S.  Liu$^{1,5}$, G. L.  Long$^{1,2,3,4,5}$,
D. M.  Tong$^5$, F. Li$^6$}
\address{
\small $^1$  Department of Physics, Tsinghua University, Beijing, 100084, China\\
$^2$ Key Laboratory For Quantum Information and Measurement, Beijing, 100084, China\\         
$^{3}$ Institute of Theoretical Physics, Chinese Academy of Sciences, Beijing 100080, P.  
R.  China\\
$^4$ Center for Atomic, Molecular and NanoSciences, \\
Tsinghua University, Beijing 100084, P. R.  China. \\
$^5$Department of Physics, Shandong Normal University,
 Jinan, 250014, China\\
 $^6$ Basic Education Section, Capital University of Economics and Business, Beijing 
100026, China}

\date{today}
\begin{abstract}
Dense coding or super-dense coding  in the case of high-dimension quantum states between 
two parties and multi-parties has been studied in this paper. We construct explicitly the 
measurement basis and the forms of the single-body unitary operations corresponding to  
the basis chosen, and the rules for selecting the one-body unitary operations in a 
multi-party case.
\end{abstract}
\pacs{03.67.-a,  89.70.+c\\
Published in Phys. Rev. A65 (2002) 022304} \maketitle

 Quantum dense-coding or
super-dense coding\cite{r1} is  one of  the important branches
of quantum information theory. It has been widely studied both in theory and in 
experiment\cite{r1,r2}.
The basic idea of quantum dense coding is  that quantum mechanics allows one to encode  
information in the quantum states that is more dense than classical coding. Bell-basis 
states
\begin{eqnarray}
|\Psi^{+}\>&=&(|00\>+|11\>)/\sqrt2,\nonumber\\
|\Psi^{-}\>&=&(|00\>-|11\>)/\sqrt2,\nonumber\\
|\phi^{+}\>&=&(|01\>+|10\>)/\sqrt2,\nonumber\\
|\phi^{-}\>&=&(|01\>-|10\>)/\sqrt2.
\end{eqnarray}
are used in dense coding.  Bell basis states are in the Hilbert space of 2 particles, each 
with 2-dimension, and they are the maximally entangled states. Supposed Alice and Bob 
share the maximally entangled state $|\Psi^{+}\>$.   Bob then operates locally on the 
particle he shares with Alice one of the four unitary transformations $I, \sigma_x, 
i\sigma_y, \sigma_z$, and this will transform $|\Psi^{+}\>$ into $|\Psi^{+}\>, 
|\phi^{+}\>, |\phi^{-}\>$ and $|\Psi^{-}\>$ respectively. Bob sends his particle back to 
Alice. Because the four manipulations result in  four orthogonal Bell states, four 
distinguishable messages, i.e., 2 bits of information then can be obtained by Alice via 
collective-measurement on the two particles. The scheme has been experimentally 
demonstrated by Mattle et. al.\cite{r3}.

With the realization of preparing high-dimension  quantum state \cite {r4}, it is of 
practical importance to study the high-dimensional aspects of various topics in quantum 
information. For example, multi-particle high-dimensional quantum teleportation has been 
constructed recently\cite{r6}. Teleportation and quantum dense coding are closely related. 
In this paper, we presents a quantum dense coding scheme between multi-parties in an 
arbitrary high dimensional Hilbert space.  As two-party dense coding is of primary 
importance, we first present the two-party dense coding scheme in arbitrary high 
dimension. Then we present the general  scheme for dense coding between multi-parties 
using high dimensional state.

To present our scheme clearer. Let's first begin with dense coding between two parties in 
3-dimension. The general Bell-basis of the Hilbert space of two particles with 3-dimension 
are \cite{r5,r6}:
\begin{equation}
|\Psi_{nm}\>=\sum\limits_je^{2\pi ijn/3}|j\>\otimes|j+m\, mod\, 3\>/\sqrt3,
\label{base2p3d}
\end{equation}
where $n$, $m$, $j=$0, 1, 2. Explicitly,
\begin{eqnarray}
|\Psi_{00}\>&=&(|00\>+|11\>+|22\>)/\sqrt3,\nonumber\\
|\Psi_{10}\>&=&(|00\>+e^{2\pi i/3}|11\>+e^{4\pi i/3}|22\>)/\sqrt3,\nonumber\\
|\Psi_{20}\>&=&(|00\>+e^{4\pi i/3}|11\>+e^{2\pi i/3}|22\>)/\sqrt3,\nonumber\\
|\Psi_{01}\>&=&(|01\>+|12\>+|20\>)/\sqrt3,\nonumber\\
|\Psi_{11}\>&=&(|01\>+e^{2\pi i/3}|12\>+e^{4\pi i/3}|20\>)/\sqrt3,\nonumber\\
|\Psi_{21}\>&=&(|01\>+e^{4\pi i/3}|12\>+e^{2\pi i/3}|20\>)/\sqrt3,\nonumber\\
|\Psi_{02}\>&=&(|02\>+|10\>+|21\>)/\sqrt3,\nonumber\\
|\Psi_{12}\>&=&(|02\>+e^{2\pi i/3}|10\>+e^{4\pi i/3}|21\>)/\sqrt3,\nonumber\\
|\Psi_{22}\>&=&(|02\>+e^{4\pi i/3}|10\>+e^{2\pi i/3}|21\>)/\sqrt3.
\label{base2p3dlist}
\end{eqnarray}
Superdense coding can be done in the following way.  Supposed Alice and Bob share the 
maximally entangled state $|\Psi_{00}\>$. Through simple calculation, it can be shown that 
the single-body operators :
\begin{eqnarray}
U_{00}&=&\left[\begin{array}{ccc}1&0&0\\ 0&1&0\\ 0&0&1\end{array}\right],\nonumber\\
U_{10}&=&
\left[\begin{array}{ccc}1&0&0\\ 0&e^{2\pi i/3 }&0\\ 0&0&e^{4\pi i/3} 
\end{array}\right],\nonumber\\
U_{20}&=&
\left[\begin{array}{ccc}1&0&0\\ 0&e^{4\pi i/3 }&0\\ 0&0&e^{2\pi i/3} 
\end{array}\right],\nonumber\\
U_{01}&=&\left[\begin{array}{ccc}0&0&1\\1&0&0\\ 0&1&0\end{array}\right],\nonumber\\
U_{11}&=&
\left[\begin{array}{ccc}0&0&e^{4\pi i/3 }\\1&0&0\\ 0&e^{2\pi i/3}&0 
\end{array}\right],\nonumber\\
U_{21}&=&
\left[\begin{array}{ccc}0&0&e^{2\pi i/3 }\\1&0&0\\ 0&e^{4\pi 
i/3}&0\end{array}\right],\nonumber\\
U_{02}&=&\left[\begin{array}{ccc}0&1&0\\ 0&0&1\\ 1&0&0\end{array}\right],\nonumber\\
U_{12}&=&
\left[\begin{array}{ccc}0&e^{2\pi i/3}&0\\ 0&0&e^{4\pi i/3}\\ 
1&0&0\end{array}\right],\nonumber\\
U_{22}&=&
\left[\begin{array}{ccc}0&e^{4\pi i/3}&0\\ 0&0&e^{2\pi i/3}\\ 1&0&0 \end{array}\right],
\label{op2p3dlist}
\end{eqnarray}
will transform $|\Psi_{00}\>$ into the corresponding states in Eq. (\ref{base2p3dlist}) 
respectively:
\begin{eqnarray}
U_{nm}|\Psi_{00}\>=|\Psi_{nm}\>.
\label{op2p3d}
\end{eqnarray}
  Bob operates one of the above unitary transformations, and sends his particle back to 
Alice. Alice takes only one measurement in the basis $ \{|\Psi_{00}\>, |\Psi_{10}\>, \dots 
, |\Psi_{22}\>\}$, and she will know what operation Bob has done, or to say, what the 
messages Bob has encoded in the quantum state.  As a result, Alice gets $\log_{2}9$ bits 
information through only one measurement. It should be pointed out that Michael Reck et. 
al.\cite{r10} have given the method to realize any discrete unitary operators and the 
operators used here can be constructed according to their protocol.

It is straightforward to generalize the above protocol to arbitrarily high dimension for 
two parties. We denote the dimension as $d$. The general Bell-basis states are
\begin{equation}
|\Psi_{nm}\>=\sum\limits_je^{2\pi ijn/d}|j\>\otimes|j+m\, mod\, d\>/\sqrt d,
\label{base2panyd}
\end{equation}
where $n$, $m$,  $j=0$,  1, $\dots$,  $d-1$. Obviously, there exist one-body operators 
$U_{nm}$ on Bob's particle that satisfy $U_{nm}|\Psi_{00}\>=|\Psi_{nm}\>$. The matrix 
elements of the unitary transformation $U_{nm}$ may be explicitly written out:
\begin{equation}
(U_{nm})_{j'j}=e^{2\pi j\;n/d}\delta_{j',j+m\; mod\; d},
\label{op2panyd}
\end{equation}
where $n$, $m$, $j=0$, 1, \dots, $d-1$. The procedure for realizing dense coding in this 
high dimension case is similar to that for the 2 dimensional case: Alice and Bob shares 
the state $|\Psi_{00}\>$, then Bob performs one of the operations in (\ref{op2panyd}) to 
his particle and sends this particle back to Alice. Alice then performs a collective 
measurement in the basis states in (\ref{base2panyd}) to find out what Bob has done to the 
particle and hence reads out the encoded message. In this case, Alice gets $\log_{2}d^{2}$ 
bits of information just making only one measurement.

Dense coding between two parties can be generalized  into {\bf multi-parties}. Bose et. 
al.\cite{r9} has generalized Bennett and Wiesner scheme of dense coding into multi-parties 
in the qubit system.    The multi-party dense coding scheme can be understood in the 
following way. There are $N+1$ users sharing an $(N+1)$-particle maximally entangled  
state, possessing one particle each. Suppose that one of them, say user 1, intends to 
receive messages from the $N$ other users. The $N$ senders  mutually decide {\it a priori} 
to perform only certain unitary operations on their particles.  After performing their 
unitary operations, each of the $N$ senders sends his particle to user 1. User 1 then 
performs a collective measurement on the $N+1$ particles and  identifies the state. Thus 
he can learn about the operation of each of the other $N$ users has done. That is to say 
that a single measurement is sufficient to reveal the messages sent by all the $N$ users. 
We now discuss the high-dimension generalization of this scenario. We also begin with an 
example with three particles in 3-dimension. We take the maximally entangle states as our 
basis:
\begin{equation}
 |\Psi_{nm}^{k}\>=\sum\limits_j e^{2\pi i\;j\;k/3}|j\>\otimes|j+n\, mod\, 3\>\otimes|j+m\, 
mod\, 3\>/\sqrt 3,
\end{equation}
where $n$, $m$, $k=0$, 1, 2. More explicitly:
\begin{eqnarray}
|\Psi_{00}^{0}\>&=&(|000\>+|111\>+|222\>)/\sqrt3, \nonumber\\
|\Psi_{01}^{0}\>&=&(|001\>+|112\>+|220\>)/\sqrt3, \nonumber\\
|\Psi_{02}^{0}\>&=&(|002\>+|110\>+|221\>)/\sqrt3, \nonumber\\
&\ldots&, \nonumber\\
|\Psi_{22}^{2}\>&=&(|022\>+e^{4\pi i/3}|100\>+e^{2\pi i/3}|211\>)/\sqrt3.
\label{base3p3dlist}
\end{eqnarray}

Suppose Alice, Bob and Claire share the maximally entangled state
\begin{equation}
|\Psi_{00}^{0}\>=(|000\>+|111\>+|222\>)/\sqrt3,
\end{equation}
and Bob and Claire hold particle 2 and 3 respectively.
The essential issue in dense coding is to find a limited number of one-body operations 
that Bob and Claire can perform so that the state $|\Psi_{00}\>$ is transformed into all 
possible states in (\ref{base3p3dlist}). Meanwhile these operations have to be able 
identified by Alice uniquely. If Bob and Claire are both allowed to perform any of the 
nine operations in (\ref{op2p3dlist}), the total number of operations are $9\times 9= 81$, 
which is greater than the total number of the basis states in (\ref{base3p3dlist}). Alice 
can not identify the operations of Bob and Claire uniquely. Thus not all the operations 
are allowed, and some restriction has to be made. In the qubit system, this problem is 
solved by allowing Bob to perform all 4 possible unitary operations $I$, $\sigma_x$, 
$i\sigma_y$ and $\sigma_z$, and Claire only performs any two of these operations.  
However, direct generalization of this rule needs care. By direct calculation, we find 
that by allowing Bob to perform all the nine operations and restricting Claire to perform 
any 3 of the nine operations will not always work. Look at table {\ref{t1}} where we have 
given the results of the operations $U_{nm}(B) U_{n'm'}(C)$ on $|\Psi_{00}\>$. Now allow 
Bob to perform all the nine unitary operations. Then from table \ref{t1} we see that the 
product of the nine operations of Bob with each operation in the subset $\{U_{00}$, 
$U_{10}$, $U_{20}\}$ of Claire can only give 9  of the 27 basis states in 
(\ref{base3p3dlist}). This is evident from table {\ref{t1}} that the first 3 columns in 
table (\ref{t1}) are just a rearrangement of the same 9 basis states. The same is true for 
the other two subsets of operations $\{U_{01}$, $U_{11}$, $U_{21}\}$ and $\{U_{02}$, 
$U_{12}$, $U_{22}\}$. Thus, Claire's 3 operations can be chosen one from each of the 3 
subsets arbitrarily. One such set is the three operations $U_{nm}$ with $n=0$: $U_{00}$, 
$U_{01}$ and $U_{02}$. Other combinations are also possible. For convenience, we can 
simply choose this subset: $\{U_{nm}$ with $n=0\}$ as the allowed operations of Claire.

With the identification of the allowed operations, say, Bob is allowed to perform all the 
nine unitary transformations, and Claire is allowed to perform the 3 operations $U_{0m}$, 
$m=0$, 1 and 2, the 3-party dense coding can be done easily. Bob and Claire perform their 
operations on their respective particles and return the particles to Alice. When Alice 
receives the operated particles from Bob and Claire, she can find what messages Bob and 
Claire have encoded by just one measurement in the basis of Eq. (\ref{base3p3dlist}) . 
That's,  Alice gets $\log_{2}27$ bits information with a single measurement.

From the above example, we can generalize the scheme into multi-party super-dense coding 
in high dimensions.  Suppose the dimension is $d$. We construct the following basis,
\begin{equation}
|\Psi_{i_{1}, i_{2}, \dots, i_{N}}^{n}\>= \sum\limits_je^{2\pi ijn/d}|j\>\otimes|j+i_{1}\, 
mod\, d\>\otimes\dots\otimes/|j+i_{N}\, mod\, d\>\sqrt d
\label{baseanypanyd}
\end{equation}
where $n, j, i_{1}$, $ i_{2}, \dots,$ $ i_{N}=0, 1, \dots, d-1$. The $d^2$ one-body 
unitary operations can be written explicitly
\begin{equation}
(U_{nm})_{j'j}=e^{2\pi jn/dN}\delta_{j',j+m\,  mod\,  d},
\label{opanypanyd}
\end{equation}
where $j=0, 1, \dots, d-1$.
Thus, if the $N$ senders mutually decide {\it a priori}  that User 2 can perform all the 
$d^2$ one-body operations in (\ref{opanypanyd}), User 3 to User $N+1$ can only perform the 
$d$ one-body operations, $U_{0m}$, $m=0,1,\ldots,$ $d-1$, operations. Then the $N$ users 
can encode their messages by performing their allowed operations to the particles at their 
proposal and return the particles to User 1. After User 1 receives all the particles, he 
can then perform a single collective measurement of all his $N+1$ particles so that he can 
reads out the messages the $N$ users have encoded. In this way,  the $N$ users can send 
out $\log_{2}d^{N+1}$ bits messages and the receiver, user 1, gets and identifies  them  
by only one measurement. This is because the $N$ senders are allowed perform $d^2\times 
d\times \dots \times d=d^{N+1}$ different combinations of the one-body unitary operations 
on the  initial $(N+1)$-particle maximally entangled state
\begin{equation}
|\Psi_{00\dots 0}\>=(|00\dots 0\>+|11\dots1\>+\dots+|d-1, d-1\,  \dots d-1\>)/\sqrt d,
\end{equation}
while there are $d^{N+1}$ orthogonal states in the Hilbert space of N+1 particles.

In summary,  we have given general schemes for multi-party high dimensional dense coding. 
Explicit expressions for the measuring basis and the forms of the one-body unitary 
transformation operators have been constructed. In particular, the operations allowed to 
users 3 to $N$ must chosen carefully.

This work is supported by
the Major State Basic Research Developed Program Grant No.  G200077400,
China National Natural Science Foundation Grant
No.  60073009,  the Fok Ying Tung Education Foundation,  and the
Excellent Young University Teachers' Fund of Education Ministry of China.

 \begin{table}
\begin{center}
\caption{Transformation table for $U_{nm}(B)U_{n'm'}(C)\Psi_{00}$}
\label{t1}
\begin{tabular}{c|ccccccccc}
&\multicolumn{9}{c}{$U_{n'm'}(C)$}\\ \cline{2-10}
 $U_{nm}(B)$ &$U_{00}$&$U_{10}$&$U_{20}$&$U_{01}$&$U_{11}$& 
$U_{21}$&$U_{02}$&$U_{12}$&$U_{22}$\\ \cline{2-10}
$U_{00}$ & $\Psi^0_{00}$ & $\Psi^1_{00}$ & $\Psi^2_{00}$ & $\Psi^0_{01}$ & $\Psi^1_{01}$ 
&$\Psi^2_{01}$ &  $\Psi^0_{02}$ & $\Psi^1_{02}$ & $\Psi^2_{02}$\\
$U_{10}$ & $\Psi^1_{00}$ & $\Psi^2_{00}$ & $\Psi^0_{00}$ & $\Psi^1_{01}$ & $\Psi^2_{01}$ & 
$\Psi^0_{01}$ & $\Psi^1_{02}$ & $\Psi^2_{02}$ & $\Psi^0_{02}$\\
$U_{20}$ & $\Psi^2_{00}$ & $\Psi^0_{00}$ & $\Psi^1_{00}$ & $\Psi^2_{01}$ & $\Psi^0_{01}$
& $\Psi^1_{01}$ & $\Psi^2_{02}$ & $\Psi^0_{02}$ & $\Psi^1_{02}$\\
$U_{01}$ & $\Psi^0_{10}$ & $\Psi^1_{10}$ & $\Psi^2_{10}$ & $\Psi^0_{11}$ & $\Psi^1_{11}$ & 
$\Psi^2_{11}$ & $\Psi^0_{12}$ & $\Psi^1_{12}$ & $\Psi^2_{12}$\\
$U_{11}$ & $\Psi^1_{10}$ & $\Psi^2_{10}$ & $\Psi^0_{10}$ & $\Psi^1_{11}$ & $\Psi^2_{11}$ & 
$\Psi^0_{11}$ & $\Psi^1_{12}$ & $\Psi^2_{12}$ & $\Psi^0_{12}$\\
$U_{21}$ & $\Psi^2_{10}$ & $\Psi^0_{10}$ & $\Psi^1_{10}$ & $\Psi^2_{11}$ & $\Psi^0_{11}$ & 
$\Psi^1_{11}$ & $\Psi^2_{12}$ & $\Psi^0_{12}$ & $\Psi^1_{12}$\\
$U_{20}$ & $\Psi^0_{20}$ & $\Psi^1_{20}$ & $\Psi^2_{20}$ & $\Psi^0_{21}$ & $\Psi^1_{21}$ &
$\Psi^2_{21}$ & $\Psi^0_{22}$ & $\Psi^1_{22}$ & $\Psi^2_{22}$\\
$U_{12}$ & $\Psi^1_{20}$ & $\Psi^2_{20}$ & $\Psi^0_{20}$ & $\Psi^1_{21}$ &$\Psi^2_{21}$ & 
$\Psi^0_{21}$ & $\Psi^1_{22}$ &$\Psi^2_{22}$ & $\Psi^0_{22}$\\
$U_{22}$ & $\Psi^2_{20}$ & $\Psi^0_{20}$ & $\Psi^1_{20}$ & $\Psi^2_{21}$ & $\Psi^0_{21}$ & 
$\Psi^1_{21}$ & $\Psi^2_{22}$ & $\Psi^0_{22}$ & $\Psi^1_{22}$\\ \hline
\end{tabular}
\end{center}
\end{table}

 \end{document}